\documentclass[%
 preprint,
 showpacs,
 preprintnumbers,
 amsmath,amssymb,
 prb,
 longbibliography,
 lengthcheck,
]{revtex4-1}

\usepackage{natbib}
\usepackage[dvipdfm]{graphicx,color}% Include figure files
\usepackage{dcolumn}% Align table columns on decimal point
\usepackage{bm}% bold math
\usepackage{hyperref}% add hypertext capabilities
\begin{document}

\title{Chirality in spin-$\frac{1}{2}$ zigzag $XY$ chain: Low-temperature density-matrix renormalization group study}

\author{Takanori Sugimoto}
\email{takanori@yukawa.kyoto-u.ac.jp}
\author{Shigetoshi Sota}
\author{Takami Tohyama}
\affiliation{Yukawa Institute for Theoretical Physics, Kyoto University, 606-8502, Kyoto, Japan}
\date{\today}

\begin{abstract}
We study spin chirality for a spin-$\frac{1}{2}$ zigzag $XY$ chain at low temperature by applying a low-temperature density-matrix renormalization group technique.  We calculate temperature dependence of dynamical and static correlations of the chirality. In a chiral phase, chiral long-range order at zero temperature disappears at finite temperature, consistent with the fact that there is no long-range order at finite temperature in one-dimensional systems with short-range interactions. In a dimer phase next to the chiral phase, we find an enhancement of static chiral correlation as well as spin correlation with increasing temperature. The enhancement corresponds to the increase in spectral weight inside a gap in the dynamical chiral-correlation function. This temperature-induced chiral correlation is a demonstration of the presence of a chiral state in excited states.
\end{abstract}


\maketitle

%%%%%%%%%%%%%%%%%%%%%%%%
\section{Introduction}
Recently low-dimensional quantum spin systems have been attracting much attention, since there are many interesting phenomena associated with magnetic frustration. A spin-$\frac{1}{2}$ zigzag $XXZ$ chain is a typical one-dimensional (1D) quantum spin system with frustration caused by competition between the first and second nearest-neighbor exchange interactions. Recent studies have shown that a vector chiral long-range order (LRO) (Ref.~1) occurs under certain conditions.~\cite{nersesyan_incommensurate_1998,lecheminant_phase_2001,hikihara_ground-state_2001}  More detailed studies of the model have been done very recently by searching for the vector chiral order in the whole parameter region.~\cite{furukawa_unpublished} It has also been shown that a magnetic field induces the vector chiral order even for a spin-$\frac{1}{2}$ zigzag Heisenberg chain.~\cite{kolezhuk_field-induced_2005,okunishi_calculation_2008,kolezhuk_field-controlled_2009} Therefore, the vector chirality can be one of important concepts characterizing 1D spin-$\frac{1}{2}$ frustrated magnets. The vector spin chirality is believed to be crucial for understanding the physics of 1D multiferroic cuprates~\cite{review,furukawa_quantum_2008} such as $\mathrm{LiCu_2O_2}$ (Refs.~11-13) and $\mathrm{LiCuVO_4}$ (Refs.~14-16).
%~\cite{masuda_competition_2004, park_ferroelectricity_2007, seki_correlation_2008}
%~\cite{naito_ferroelectric_2007, yasui_relationship_2008, schrettle_switching_2008}

The ground-state properties of the spin-$\frac{1}{2}$ zigzag $XXZ$ chain have extensively examined numerically~\cite{hikihara_ground-state_2001,furukawa_unpublished} and analytically.~\cite{batista_canted_2009} However, physical properties at finite temperatures have not been studied as far as we know. Even for a simpler case without anisotropic $z$ component, i.e., zigzag $XY$ chain, there is no work on temperature dependence of physical quantities.  In particular, it is interesting to clarify how the chiral properties change at finite temperature.

In the chiral phase of the spin-$\frac{1}{2}$ zigzag $XXZ$ chain, it is known that time-reversal symmetry and global $U$(1) symmetry are preserved but parity symmetry is broken. It is interesting to clarify how finite temperature affects on such a chiral phase and chiral properties in the model.

In this paper, we investigate finite-temperature properties of the spin-$\frac{1}{2}$ zigzag $XY$ chain that is the simplest frustrated quantum spin model. The model shows vector chiral LRO at zero temperature in some parameter regions.~\cite{okunishi_calculation_2008} It is known that the chiral ordered phase is located next to a dimer phase.~\cite{hikihara_ground-state_2001} In order to clarify the effect of temperature on the chirality in both the chiral and dimer phases, we employ a recently developed low-temperature density matrix renormalization group (LT-DMRG) technique~\cite{sota_low-temperature_2008} and calculate temperature dependence of dynamical and static correlations of the chirality. In the chiral phase, we find that the chiral LRO disappears at finite temperature as expected form the fact that there is no LRO at finite temperature in 1D systems with short-range interactions. In the dimer phase, static chiral correlation as well as spin correlation is enhanced with increasing temperature, followed by subsequent suppression with further increase in temperature. The maximum of the correlations appears at a temperature where spectral weight inside a gap in the dynamical chiral correlation function increases. This temperature-induced chiral correlation is a demonstration of the presence of a chiral state in excited states.

This paper is organized as follows. The spin-$\frac{1}{2}$ zigzag $XY$ model is introduced in Sec.~II. In Sec.~III, LT-DMRG is briefly explained and quantities examined are introduced. The dynamical chiral correlation functions at both zero and finite temperatures are shown in Sec.~IV.~A. The static correlation functions are presented in Sec.~IV.~B. Conclusions are summarized in Sec.~V.

%%%%%%%%%%%%%%%%%%%%
\section{Model}
The Hamiltonian of the spin-$\frac{1}{2}$ zigzag $XY$ chain is given by
\begin{equation}
\mathcal{H}=\sum_{\rho=1}^2 J_{\rho} \sum_{l} \left( S_l^xS_{l+\rho}^x + S_{l}^yS_{l+\rho}^y \right),
\label{eq:hamiltonian}
\end{equation}
where $S_l^{x(y)}$ is the $x(y)$ component of spin operator at site $l$, and $J_{\rho}$ is the nearest ($\rho=1$) and next nearest ($\rho=2$) neighboring spin coupling constant. In the present paper, we take antiferromagnetic spin couplings, i.e., $J_1>0$ and $J_2>0$, which lead to a frustrated spin system. At zero temperature, three phases have been identified:~\cite{okunishi_calculation_2008} gapless chiral phase ($\alpha_J \equiv J_1/J_2 \lesssim 0.81$), gapped dimer phase ($0.81 \lesssim \alpha_J \lesssim 3.1$), and Tomonaga-Luttinger liquid phase ($\alpha_J \lesssim 3.1$).  In the chiral phase, vector chirality 
\begin{equation}
\kappa_l = S_l^x S_{l+1}^y - S_l^yS_{l+1}^x
\label{eq:op_chirality}
\end{equation}
shows a finite expectation value that is evidence of LRO.

%%%%%%%%%%%%%%%%%%%
\section{Method}
In the present study, we examine effects of both quantum and thermal fluctuations on the frustrated spin-$\frac{1}{2}$ zigzag $XY$ chain.  However, in such a frustrated system, it is difficult to employ standard Jordan-Wigner transformation, because phase factor corresponding to a next-nearest neighbor interaction remains after the transformation.  Thus, we use the DMRG method~\cite{white_density-matrix_1992} to analyze properties of this frustrated system.  

In the standard DMRG method at zero temperature, we consider the ground state as a target state.  At finite temperatures, on the other hand, we employ a state $|\tilde{\xi}\rangle$ as a target state.  The state $|\tilde{\xi}\rangle$ is defined as a product of the Boltzmann factor and a normalized state $|\xi\rangle$ randomly generated,~\cite{sota_low-temperature_2008} namely,
\begin{equation}
|\tilde{\xi}\rangle = e^{-\beta \mathcal{H}/2} |\xi \rangle,
\end{equation}
where $\beta=1/T$ with temperature $T$. (Hereafter we take the Boltzmann constant $k_{\rm B}=1$ and $\hbar=1$.)  Since an arbitrary state can be expanded with a complete set of basis $|\zeta_i\rangle$, the target state $|\tilde{\xi}\rangle$ is rewritten by
\begin{equation}
|\tilde{\xi}\rangle = e^{-\beta \mathcal{H}/2} \sum_{i=1}^N r_i|\zeta_i\rangle,
\end{equation}
where $N$ is the total number of the basis and $r_i=\langle \zeta_i | \xi \rangle$ is a normalized random number.  In our procedure, $|\zeta_i\rangle$ is constructed by the basis of a total system in DMRG procedure called superblock.  $|\tilde{\xi}\rangle$ is obtained by the regulated polynomial expansion of the Boltzmann factor.\cite{sota_low-temperature_2008}

We next explain how to calculate thermodynamical quantities.  $|\zeta_i\rangle$ can also be expanded by eigenstate $|\epsilon_n\rangle$ with eigenenergy $\epsilon_n$.  
$|\tilde{\xi}\rangle $ thus reads
\begin{equation}
|\tilde{\xi}\rangle = \sum_{i=1}^N \sum_{n=1}^N r_i b_{i,n} e^{-\beta \epsilon_n/2}  |\epsilon_n\rangle \label{eq:expnd}
\end{equation}
with $b_{i,n}=\langle \epsilon_n | \zeta_i \rangle$.  We consider a normalized expectation value of an arbitrary operator $A$ in terms of $|\tilde{\xi}\rangle$,
\begin{equation}
\langle\langle A\rangle\rangle \equiv \frac{\langle \tilde{\xi} | A | \tilde{\xi}\rangle}{\langle \tilde{\xi} | \tilde{\xi}\rangle} \label{eq:llArr}.
\end{equation}
By using Eq.~(\ref{eq:expnd}), the denominator is given by 
\begin{equation}
\langle\tilde{\xi}|\tilde{\xi}\rangle = \sum_n e^{-\beta \epsilon_n} \left[\sum_{i}r_i^2b_{i,n}^2+\sum_{i\neq j}r_ir_jb_{i,n}b_{j,n}\right]. \label{eq:xixi}
\end{equation}
If the coefficient $r_i$ is randomly generated from a rectangular distribution and a plenty of the random number set is averaged, a square of $r_i$ converges on a non-zero constant $\langle r_i^2\rangle_{r} = C$, where $ \langle \cdots \rangle_r$ denotes random sampling and averaging.  A product $\langle r_ir_j \rangle_{r}$ converges on zero because positive and negative values appear with same probability during averaging.  After random sampling and averaging, the denominator is expected to be a product of the constant $C$ and the partition function $Z$, 
\begin{equation}
\langle\tilde{\xi}|\tilde{\xi}\rangle = C \sum_n e^{-\beta \epsilon_n} = C Z,
\label{eq:xixiCZ}
\end{equation}
because of the normalized condition $\sum_ib_{i,n}^2=1$.
The numerator of $\langle\langle A \rangle\rangle$ in Eq.~(\ref{eq:llArr}) reads 
\begin{equation}
\langle \tilde{\xi} | A | \tilde{\xi}\rangle = \sum_n S_{n,n} + \sum_{n\neq m} S_{n,m},
\end{equation}
where
\begin{equation}
S_{n,n} = e^{-\beta \epsilon_n} \langle \epsilon_n | A | \epsilon_n \rangle  \left[\sum_{i}r_i^2b_{i,n}^2+\sum_{i\neq j}r_ir_jb_{i,n}b_{j,n}\right] \label{eq:Snn}
\end{equation}
and
\begin{eqnarray}
S_{n,m} &=& e^{-\beta (\epsilon_n + \epsilon_m)/2} \langle \epsilon_n | A | \epsilon_m \rangle \nonumber \\
 && \times \left[\sum_{i}r_i^2b_{i,n}b_{i,m}+\sum_{i\neq j}r_ir_jb_{i,n}b_{j,m}\right]. \label{eq:Snm}
\end{eqnarray}
Similar to the denominator, after random sampling and averaging, the bracket $[\cdots ]$ in Eq.~(\ref{eq:Snn}) converges on the constant $C$.  On the other hand, the bracket in Eq.~(\ref{eq:Snm}) converges on zero because $\langle r_i^2 \rangle_{r} = C$, $\langle r_ir_j \rangle_{r} = 0$, and $\sum_i b_{i,n}b_{i,m}=\delta_{n,m}$.  The numerator of $\langle\langle A \rangle\rangle$ then reads 
\begin{equation}
\langle\tilde{\xi} | A | \tilde{\xi}\rangle = C \sum_n e^{-\beta \epsilon_n} \langle \epsilon_n | A | \epsilon_n \rangle. \label{eq:Aav}
\end{equation}
From Eqs.~(\ref{eq:xixiCZ}) and (\ref{eq:Aav}), $\langle\langle A \rangle\rangle$ becomes the same as the thermodynamical average
\begin{equation}
\langle A \rangle = \frac{1}{Z}\sum_n e^{-\beta \epsilon_n}\langle \epsilon_n | A | \epsilon_n \rangle.
\end{equation}

Dynamical correlation function for an arbitrary operator $A$ is given as
\begin{equation}
\chi(\omega)=-\frac{1}{\pi LZ}\sum_n e^{-\beta \epsilon_n} \Im \langle \epsilon_n| A^\dagger \frac{1}{\omega - \mathcal{H} + \epsilon_n +i\gamma} A |\epsilon_n\rangle,
\end{equation}
where $L$ is the system size and $\gamma$ is an infinitesimally small energy. In our DMRG simulations, we take $\gamma$ to be a small but finite value, which makes a delta function to a Lorentzian-type broaden function.
To calculate the dynamical correlation function, we introduce an integral
\begin{equation}
\chi^\prime(\omega) = -\frac{1}{\pi L \langle \tilde{\xi}|\tilde{\xi}\rangle} \int_{-\infty}^{\infty} d\epsilon e^{-\beta \epsilon/2} \Im \langle \tilde{\xi} | A^\dagger \frac{1}{\omega - \mathcal{H} + \epsilon +i\gamma} A |\epsilon \rangle, \label{eq:chi_prime}
\end{equation}
where $|\epsilon \rangle=\delta(\epsilon-\mathcal{H})|\xi\rangle$.  The integral $\chi^\prime(\omega)$ is expected to converge to $\chi(\omega)$ after random sampling and averaging mentioned above.\cite{sota_low-temperature_2008}  

During DMRG procedure before the final calculation of Eq.~(\ref{eq:chi_prime}), we make use of the following approximation to prepare a target state for dynamical quantity:
\begin{equation}
\int_{-\infty}^{\infty} d\epsilon e^{-\beta \epsilon/2} \frac{1}{\omega-\mathcal{H}+\epsilon+i\gamma}A|\epsilon \rangle \simeq \frac{1}{\omega-\mathcal{H}+\tilde{E}+i\gamma}A|\tilde{\xi}\rangle,
\end{equation} 
where $\tilde{E}\equiv \langle\tilde{\xi}| \mathcal{H} |\tilde{\xi} \rangle$.\cite{sota_low-temperature_2008}  

In dynamical DMRG calculations, we use three target states: $|\epsilon_0\rangle$, $A|\epsilon_0\rangle$, and $[\omega-\mathcal{H}+\epsilon_0+i\gamma]^{-1}A|\epsilon_0\rangle$ at zero temperature, and $|\tilde{\xi}\rangle$, $A|\tilde{\xi}\rangle$, and $[\omega-\mathcal{H}+\tilde{E}+i\gamma]^{-1}A|\tilde{\xi}\rangle$ at finite temperatures.

To study the chiral order, we calculate a chirality correlation function, which is defined by
\begin{equation}
C_{\kappa}(r_l) = \frac{1}{S^4} \langle \kappa_{l_L} \kappa_{l_R} \rangle \label{eq:Ckr},
\end{equation}
where $S=1/2$ and $l_L$ ($l_R$) denotes the left (right) site of chirality,
\begin{equation}
l_L = \frac{L}{2} - \mathrm{int} \left( \frac{r_l}{2} \right)
\end{equation}
and
\begin{equation}
l_R = \frac{L}{2} + \mathrm{int} \left( \frac{r_l+1}{2} \right)
\end{equation}
with $r_l$ being the distance between both sites. The integer part of $x$ is represented by $\mathrm{int}(x)$. 

We also examine spin-correlation function, whose momentum representation is assumed to be
\begin{equation}
C_{S}(q) = \frac{1}{LS^2}\sum_{r_l} \cos(qr_l) \langle S_{l_L}^x S_{l_R}^x \rangle \label{eq:CSq}.
\end{equation}

Dynamical chiral-correlation function is defined by
\begin{eqnarray}
K(q,\omega)&=&-\frac{1}{\pi LZS^4} \sum_{n=1}^N e^{-\beta \epsilon_n} \nonumber \\
&& \times \Im \langle \epsilon_n| \kappa(q) \frac{1}{\omega - \mathcal{H} + \epsilon_n +i\gamma} \kappa(q) |\epsilon_n\rangle, \label{eq:chrl_dcf}
\end{eqnarray}
where $\kappa(q)$ is the momentum representation of chirality under open boundary condition:
\begin{equation}
\kappa(q) = \sqrt{\frac{2}{L+1}} \sum_l \sin(ql) \kappa_l
\end{equation}
with
\begin{equation}
q=\frac{n\pi}{L+1}\> \left(n =1,2,\dots ,L \right).
\end{equation}
We note that open boundary condition is better in getting a good convergence than periodic boundary condition~\cite{white_density-matrix_1992}

%%%%%%%%%%%%%%%%%%%%%%
\section{Results}
\subsection{Dynamical chiral-correlation function}
\begin{figure}[tb]
\includegraphics[scale=0.45]{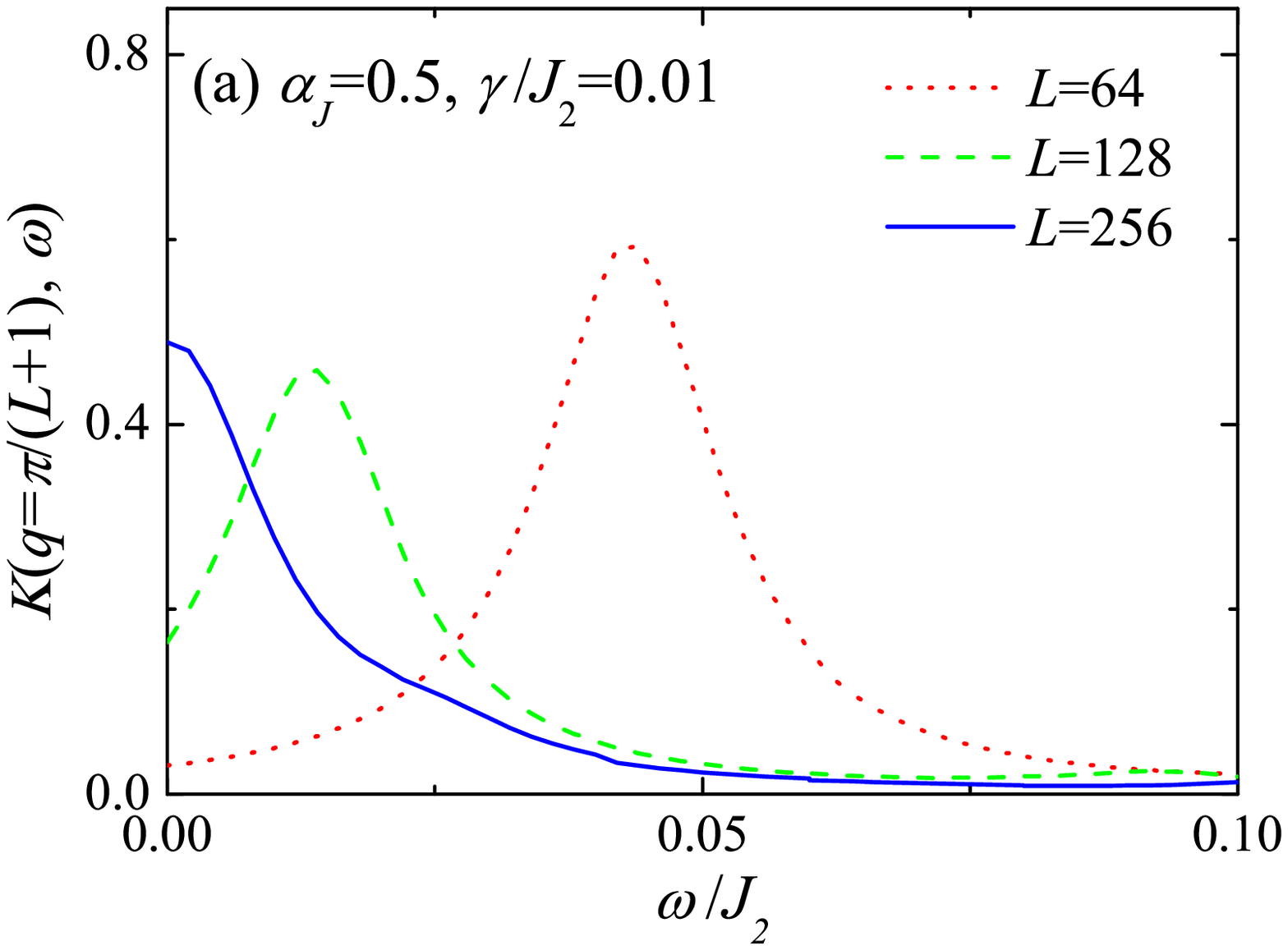}
\includegraphics[scale=0.45]{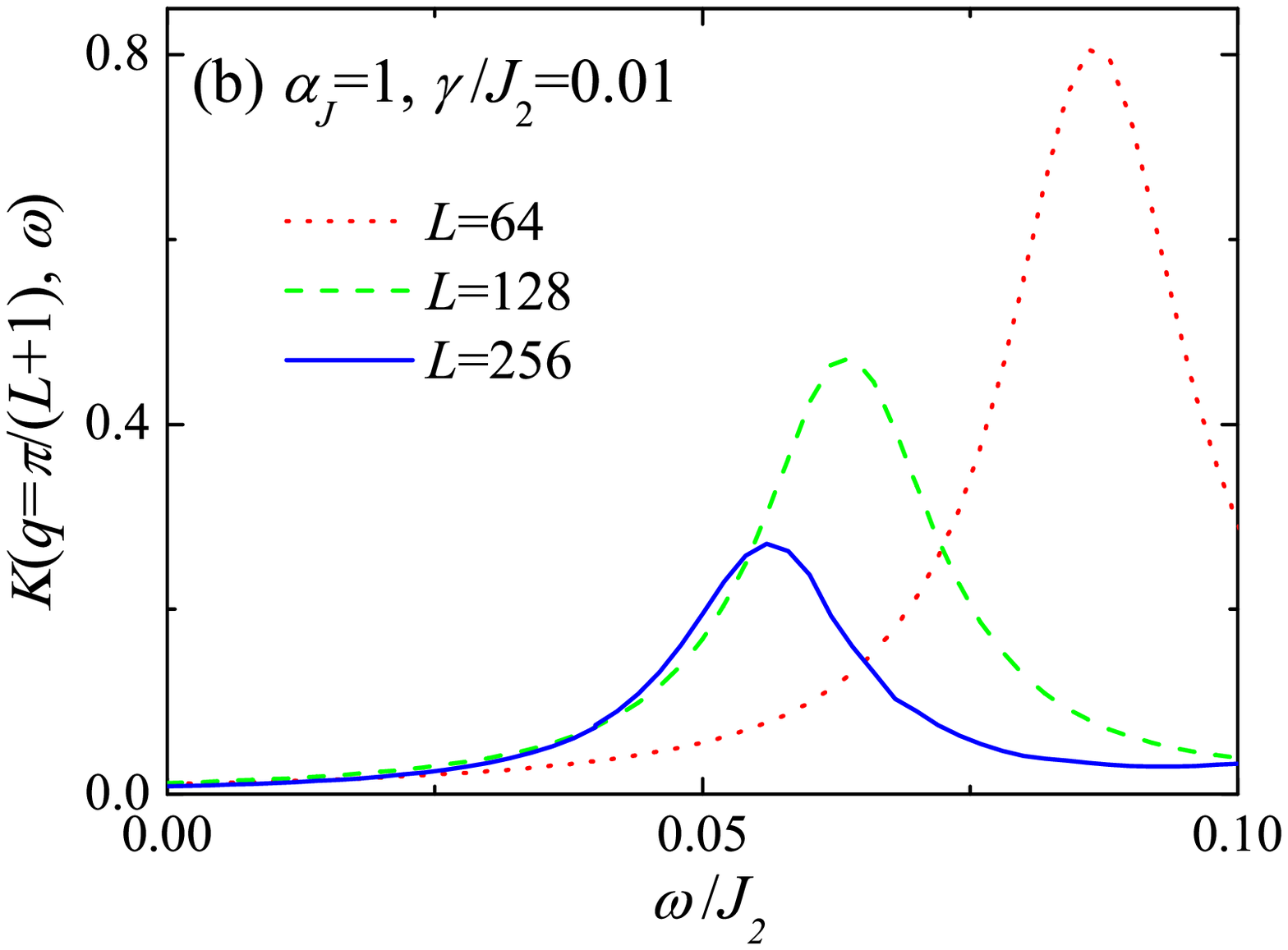}
\caption{\label{fig:chrl_dcf_t0} (Color online) Dynamical chiral correlation function $K(q,\omega)$ in spin-$\frac{1}{2}$ zigzag $XY$ chains with various system size $L$ at zero temperature. (a) $\alpha_J=0.5$ and (b) $\alpha_J=1$. The broadening factor $\gamma=0.01J_2$. The momentum $q$ is the smallest one defined in each system. The maximum DMRG truncation number is $m=400$ for $L=256$.}
\end{figure}

\begin{figure}[tb]
\includegraphics[scale=0.45]{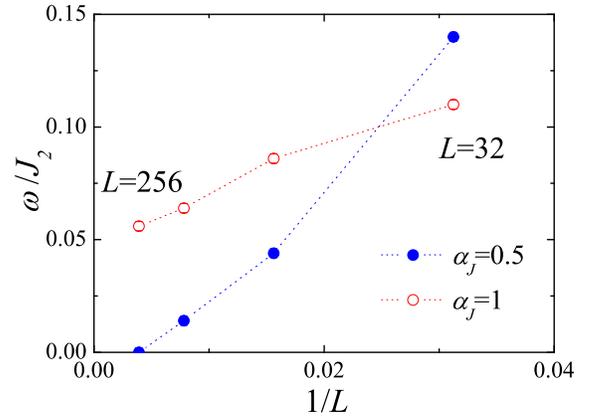}
\caption{\label{fig:chrl_dcf_t0_size} Peak position of the spectra in Fig.~\ref{fig:chrl_dcf_t0} as a function of the inverse system size. }
\end{figure}

We calculate the dynamical chiral-correlation function $K(q,\omega)$ to confirm chiral properties of the ground state at $\alpha_J=0.5$ (the chiral phase) and $\alpha_J=1.0$ (the dimer phase). 
Figure~\ref{fig:chrl_dcf_t0}(a) shows $K(q,\omega)$ in various sizes in the chiral phase ($\alpha_J=0.5$). 
The momentum $q$ is taken to be the smallest one on each system, i.e., $q=\pi/(L+1)$.
The factor $\gamma$ in Eq.~(\ref{eq:chrl_dcf}), which makes a delta function Lorentzian shape, is taken to be $0.01J_2$. Since a peak structure in $K(q,\omega)$ is well described by a single Lorentzian with the half width at half maximum, $\gamma$, its peak position is regarded as the position of the delta function. 
The peak position approaches zero energy $\omega=0$ 
with increasing the system size. The size dependence of the peak position is plotted in Fig.~\ref{fig:chrl_dcf_t0_size}. 
We find that an excited state forming the peak degenerates with the ground state in the thermodynamic limit $L \rightarrow \infty$. 
This is simply due to the presence of degeneracy in the ground state as a result of chiral LRO. 

In Ref.~10, excited states above the peak position have been examined by the exact diagonalization (ED) method and no gap has been identified between the ground state and the excited states. In the present calculation, it is difficult to identify precise energy of such excited states. For example, there is a very small hump structure at $\omega\sim 0.25J_2$ for $L=256$ in Fig.~\ref{fig:chrl_dcf_t0}(a) but precise determination of the energy of a corresponding eigenstate is impossible unlike the ED method.~\cite{furukawa_quantum_2008}

Peak positions for $\alpha_J=1.0$ (the dimer phase) in Fig.~\ref{fig:chrl_dcf_t0}(b) approach a finite energy with increasing the system size accompanied by the reduction in spectral weight. Figure~\ref{fig:chrl_dcf_t0_size} shows that gap magnitude in the thermodynamic limit is $\omega \sim 0.05 J_2$. The spectral weight at the gap energy seems to vanish in the thermodynamic limit [see Fig.~ \ref{fig:chrl_dcf_t0}(b)], in contrast to the case of the chiral phase.

\begin{figure}[tb]
\includegraphics[scale=0.45]{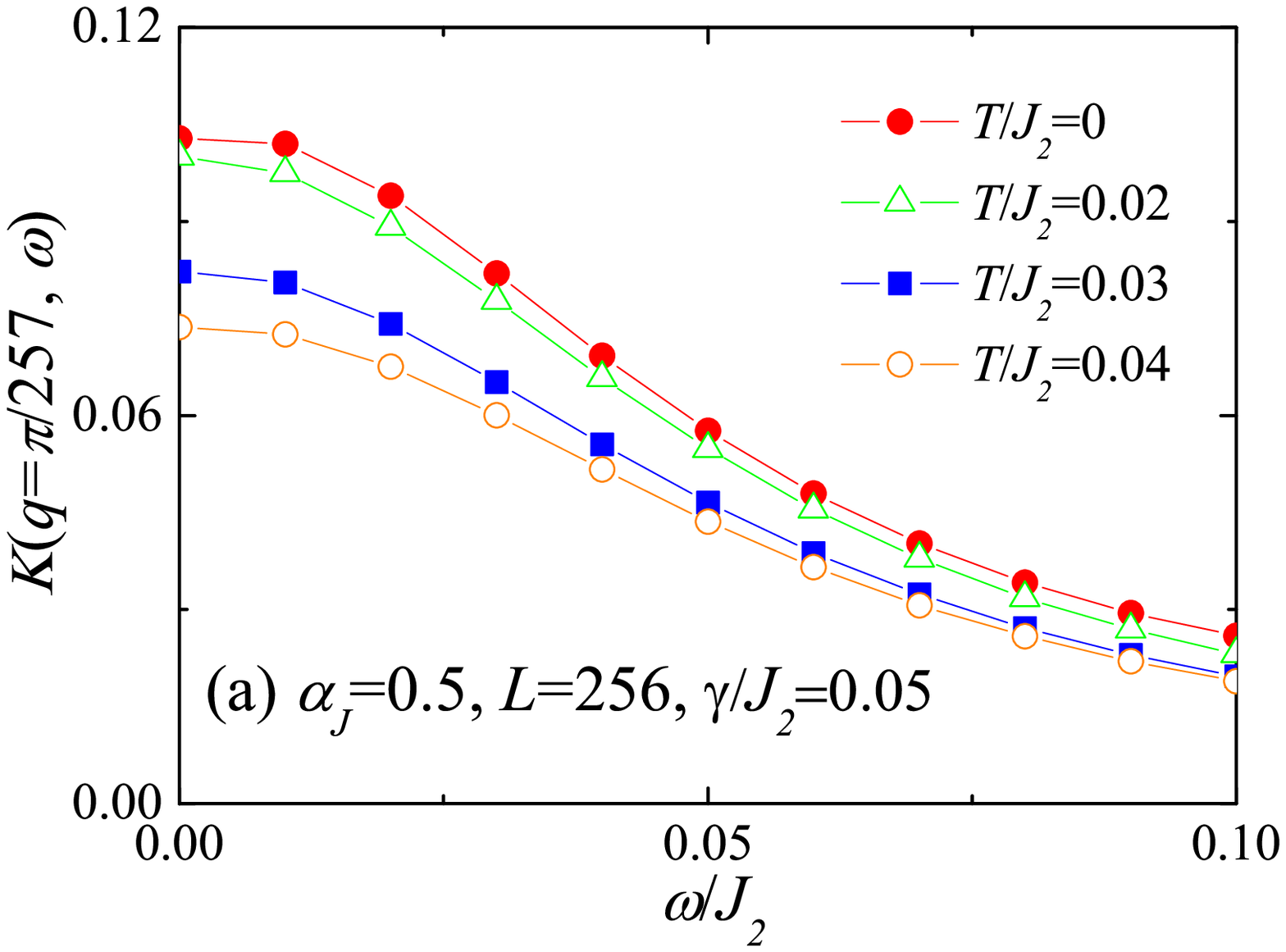}
\includegraphics[scale=0.45]{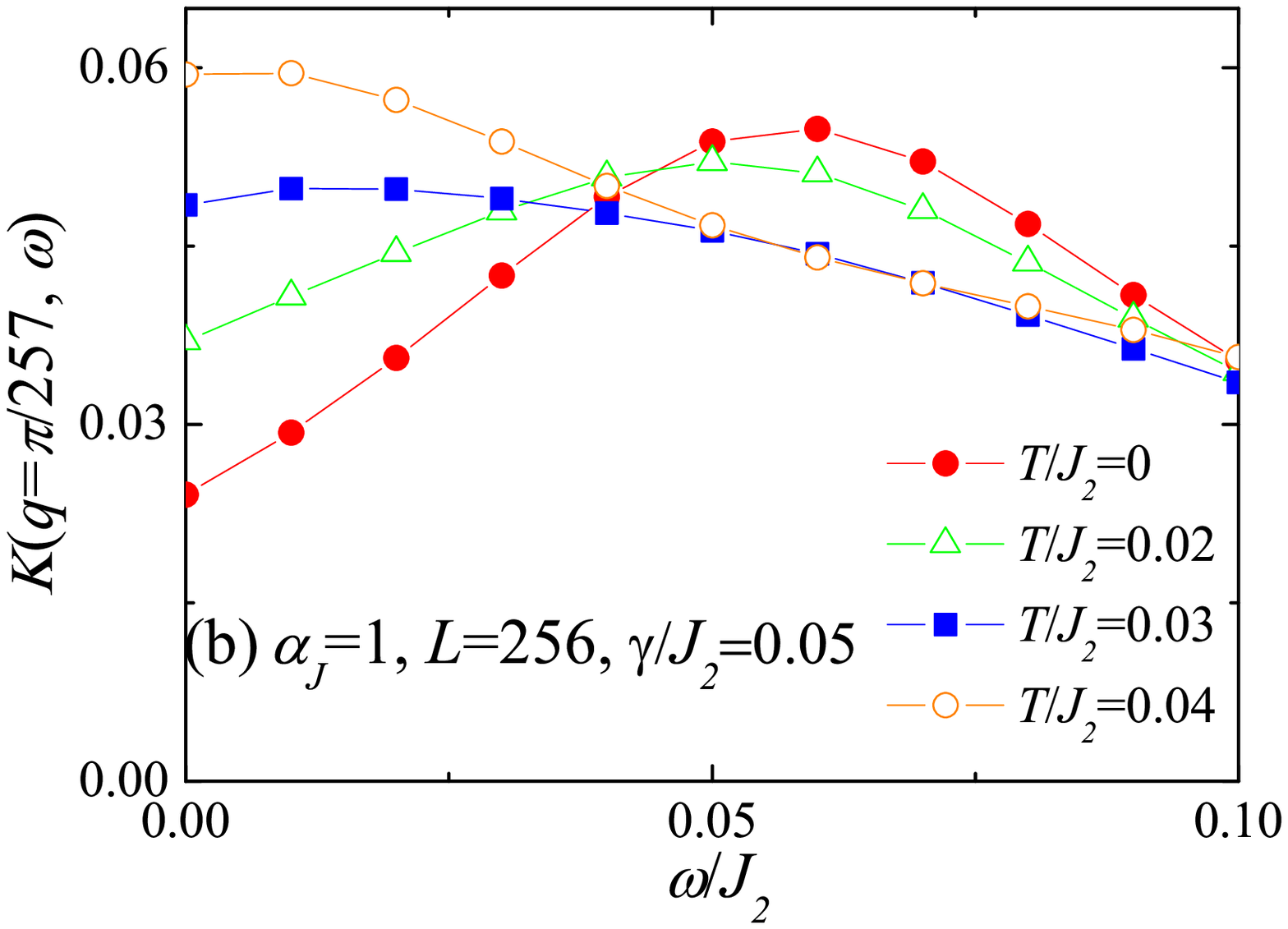}
\caption{\label{fig:chrl_dcf_ft} (Color online) Temperature dependence of dynamical chiral correlation function $K(q,\omega)$ in a 256-site spin-$\frac{1}{2}$ zigzag $XY$ chain for (a) $\alpha_J=0.5$ and (b) $\alpha_J=1$.  The broadening factor $\gamma=0.05J_2$.  The maximum DMRG truncation number is $m=500$.}
\end{figure}

In order to investigate the effect of temperature on chiral excitation, 
we calculate $K(q,\omega)$ at finite temperatures for the $L=256$ zigzag chain.
At $\alpha_J=0.5$, spectral weight decreases monotonously with increasing temperature as shown in Fig.~\ref{fig:chrl_dcf_ft}(a). 
On the contrary, $K(q,\omega)$ for the dimer phase [Fig.~\ref{fig:chrl_dcf_ft}(b)] shows a non-monotonic behavior: the peak position shifts to the lower energy side accompanied by an enhancement of spectral weight below $\omega\sim 0.05J_2$. At $T/J_2=0.04$, the $\omega$ dependence is qualitatively similar to the case of $\alpha_J=0.5$ at the same temperature. There is a crossover from gapped to gapless excitation between $T/J_2=$0.02 and 0.04.

\subsection{Static correlation functions}
\begin{figure}[tb]
\includegraphics[scale=0.65]{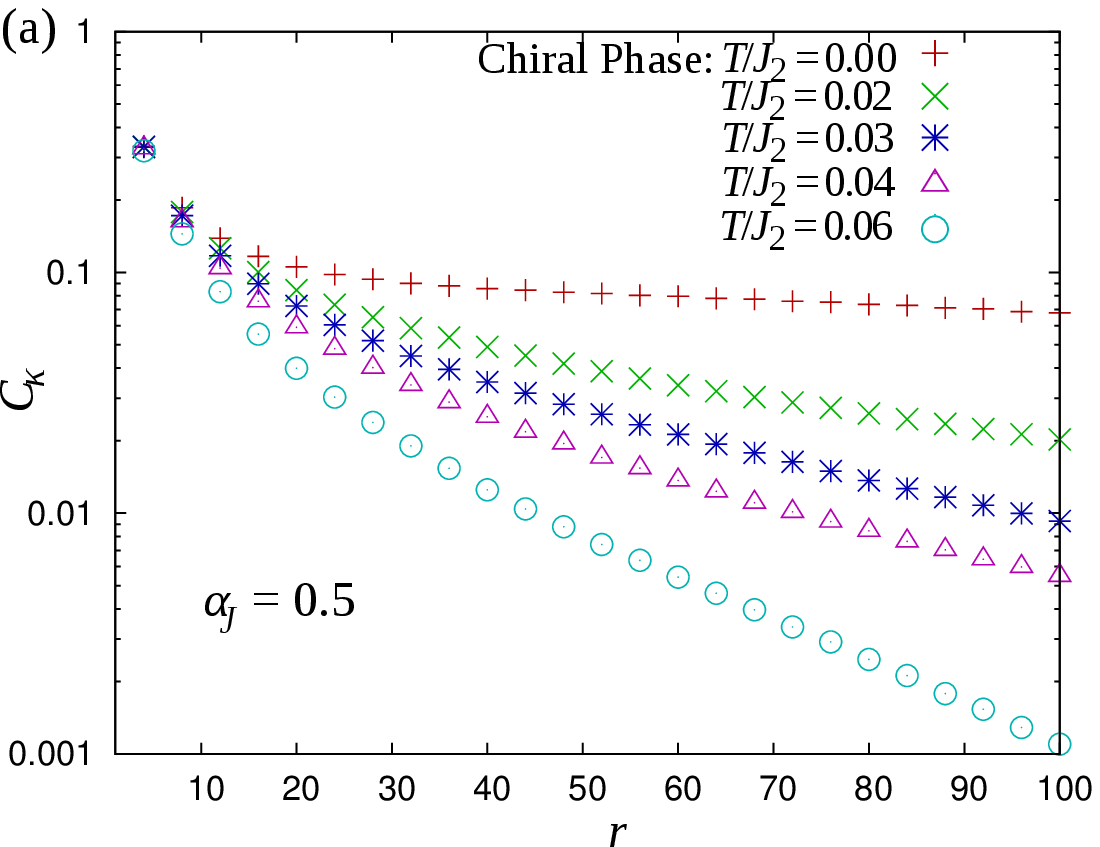}
\includegraphics[scale=0.65]{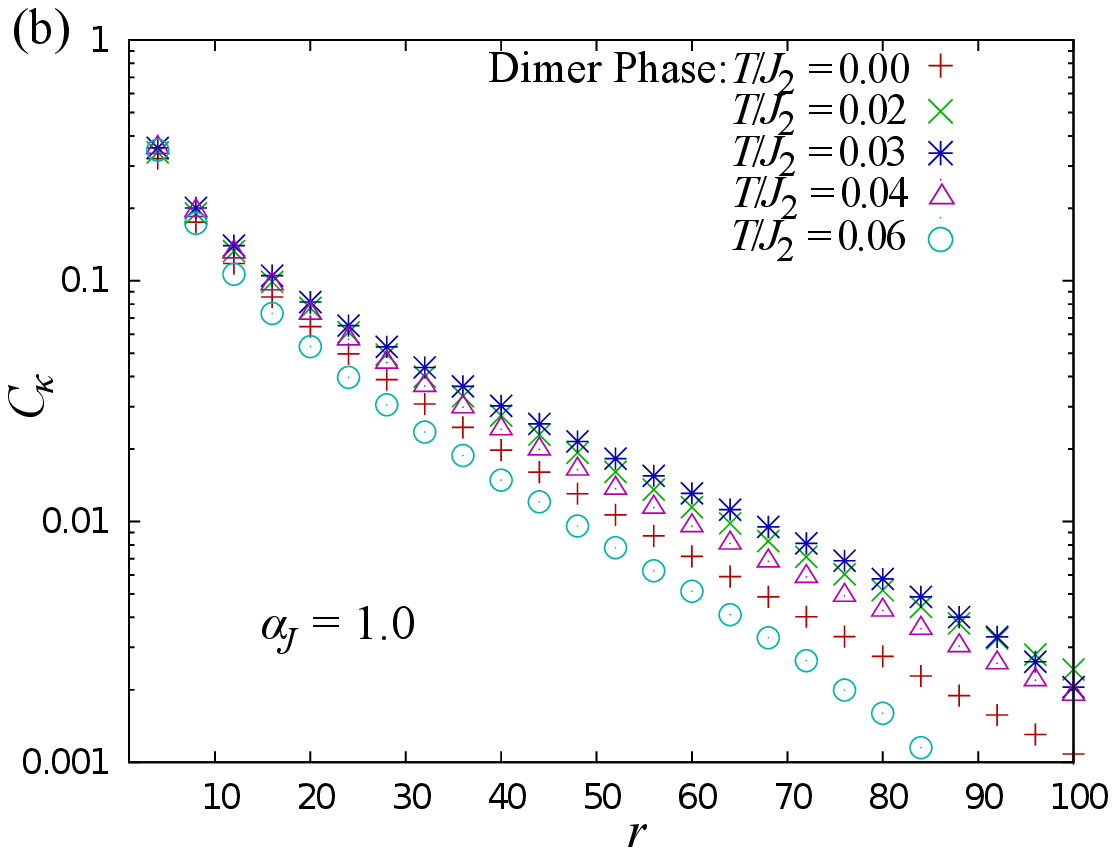}
\caption{\label{fig:chrl_scf} (Color online) Temperature dependence of chiral correlation function $C_{\kappa}(r)$ in a 256-site spin-$\frac{1}{2}$ zigzag $XY$ chain for (a) $\alpha_J=0.5$ and (b) $\alpha_J=1.0$ as a function of the distance $r$.  The maximum DMRG truncation number is $m=500$.}
\end{figure}

In order to understand the behaviors of dynamical chiral correlation mentioned above, we calculate temperature dependence of static chiral-correlation function $C_{\kappa}(r)$ as a function of distance $r$. Figure~\ref{fig:chrl_scf} shows $C_{\kappa}(r)$ up to $r=100$ for the $L=256$ system.  We found that finite-size effects seem to be serious when $r$ is more than 100, judging from the fact that the value of the correlation is independent of system size up to $r\sim20$ and $r\sim40$ for $L=64$ and $L=128$, respectively. Concerning convergence of the data in terms of the truncation number $m$ in LT-DMRG, we checked that $m=500$ gives a good convergence even for the highest temperature $T=0.06J_2$ examined.  At zero temperature for $\alpha_J=0.5$ shown in Fig.~\ref{fig:chrl_scf}(a), $C_{\kappa}(r)$ is almost constant, representing the presence of chiral LRO.~\cite{hikihara_ground-state_2001}
We note that the data show a very slight decrease beyond $r\sim80$. This is actually the finite-size effect.
In contrast to the zero-temperature result, $C_{\kappa}(r)$ decreases with $r$ at $T/J_2=0.02$, indicating that the LRO at zero temperature disappears at finite temperature.
The difference between the $T=0$ and $T=0.02 J_2$ results becomes clear when we plot the data with a log-log graph (not shown): The $T=0$ data exhibit an upward bending that is an indication of LRO (Ref.~4) while the $T=0.02 J_2$ data do not show such an upward bending.
With further increasing temperature, $C_{\kappa}(r)$ decreases monotonously. We note that the monotonous decrease in $C_{\kappa}(r)$ corresponds to the monotonous decrease in $K(q,\omega)$ with temperature as shown in Fig.~\ref{fig:chrl_dcf_ft}(a). 

We also performed the LT-DMRG calculate of an order parameter $\langle\kappa_{L/2}\rangle$ for $L=400$ chain, according to a technique proposed in Ref.~7 (not shown here).  We found that, at $T/J_1=0.01$, $\langle\kappa_{L/2}\rangle$ becomes zero. Together with the results of $C_{\kappa}(r)$ in Fig.~\ref{fig:chrl_scf}(a), we conclude that no chiral LRO at finite temperature in the chiral phase of the spin-$\frac{1}{2}$ zigzag chain.  This is consistent with the fact that there is no LRO at finite temperature in 1D systems with short-range interactions. 

Figure~\ref{fig:chrl_scf}(b) shows $C_{\kappa}(r)$ for $\alpha_J=1.0$ without chiral LRO. At $T=0$, $C_{\kappa}(r)$ rapidly decreases with $r$ as expected. With increasing temperature, $C_{\kappa}(r)$ at all $r$ increases and then decreases, showing maximum values at $T/J_2=0.03$. This temperature corresponds to the crossover temperature in $K(q,\omega)$ discussed above. The enhancement of $C_{\kappa}(r)$ can be attributed to the contribution of excited states with strong chiral correlation. It is natural to assign the eigenstate at the gap position of $K(q,\omega)$ as such an excited state. In fact, corresponding gap energy $\omega\sim 0.05J_2$ is close to the crossover temperature. Therefore, this temperature-induced chiral correlation together with the enhancement of low-energy chiral excitations in $K(q,\omega)$ is a demonstration of the presence of a chiral state in excited states in the dimer phase. 

It is known that chiral LRO at the ground state does not need LRO of spin.~\cite{hikihara_ground-state_2001}  In fact, spin-correlation function decays in the power law while chiral-correlation function remains finite in the limit of $r \rightarrow \infty$. However, the magnitude of the spin correlation seems to be correlated with that of the chiral correlation.~\cite{hikihara_ground-state_2001} Therefore, it is expected that the spin correlation shows temperature dependence similar to that of the chiral correlation. In order to confirm this, we also calculate spin correlation function $C_{S}(q)$ in Eq.~(\ref{eq:CSq}).

\begin{figure}[tb]
\includegraphics[scale=0.65]{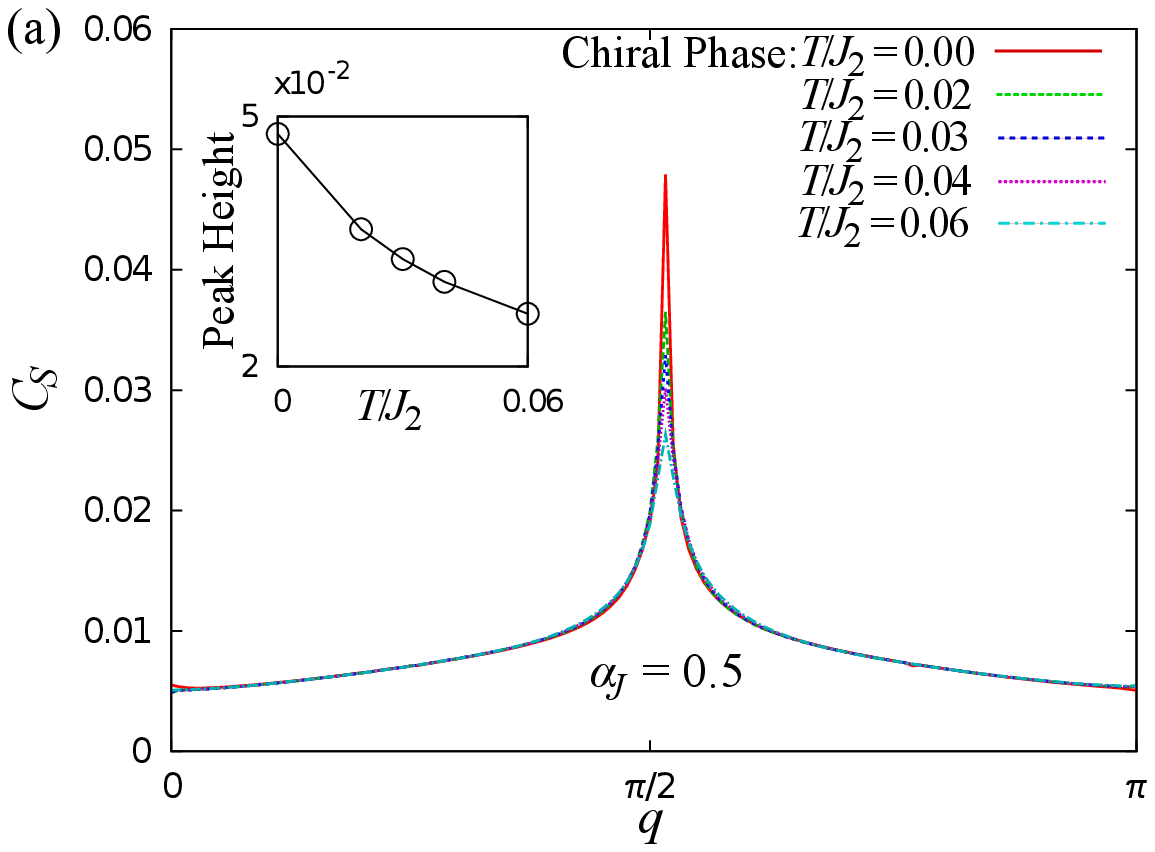}
\includegraphics[scale=0.65]{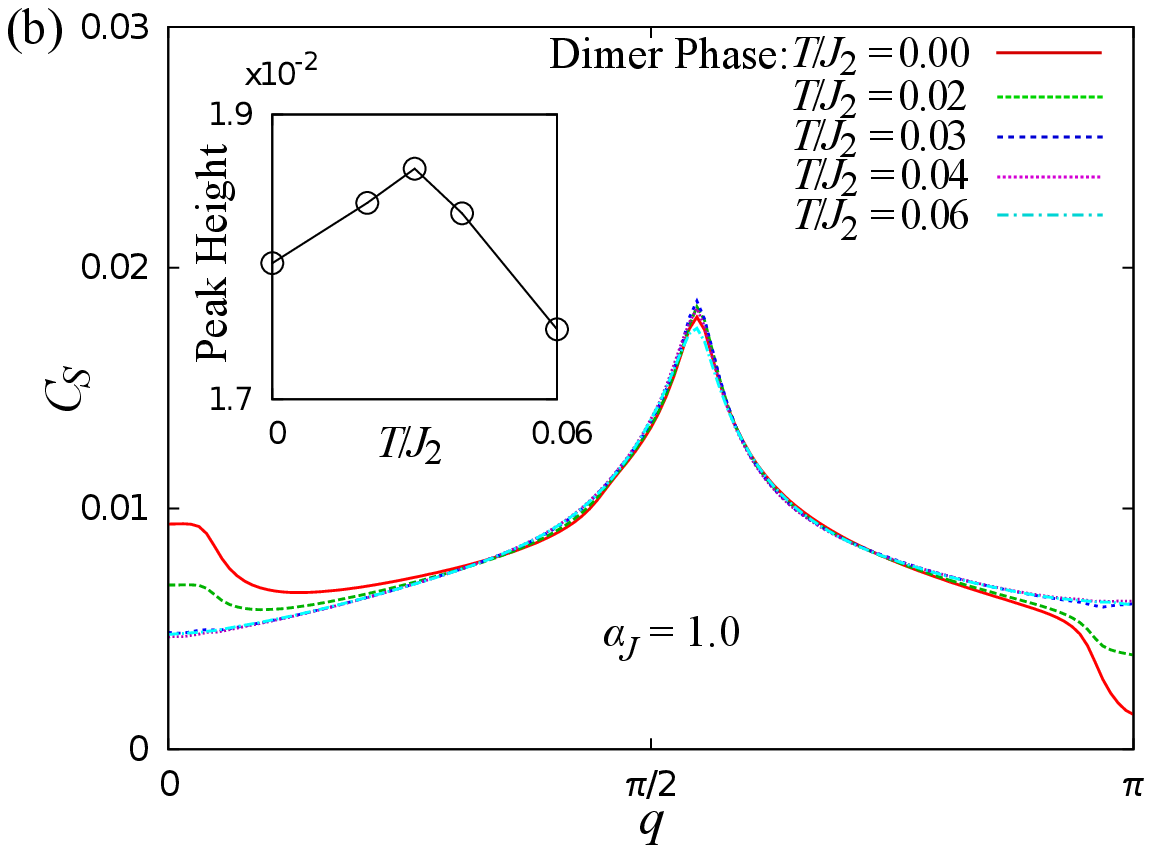}
\caption{\label{fig:spin_scf_frr} (Color online) Temperature dependence of spin correlation function $C_{S}(q)$ in a 256-site spin-$\frac{1}{2}$ zigzag $XY$ chain for (a) $\alpha_J=0.5$ and (b) $\alpha_J=1.0$ as a function of the wave number $q$. The maximum DMRG truncation number is $m=500$. Inset shows the height of the peaks as a function of temperature.}
\end{figure}

Figure~\ref{fig:spin_scf_frr} shows $C_{S}(q)$ for $\alpha_J=0.5$ and $1.0$.  For $\alpha_J=0.5$, peak height at $q\sim0.53\pi$ decreases with increasing temperature. The decrease is similar to that in the chiral correlation shown in Fig.~\ref{fig:chrl_scf}(a).
In the case of the dimer phase ($\alpha_J=1$), peak height increases up to the characteristic temperature $ T/J_2 \sim 0.03$ before it decreases, similar to the behavior of $C_{\kappa}(r)$. The correspondence between the chiral and spin correlations, thus, confirms the presence of excited states that have strong chiral correlation in the dimer phase.
The position of the peaks is not the same for $\alpha_J = 0.5$ and for $\alpha_J = 1$. The position is consistent with a characteristic vector of the spiral phase in the classical limit, given by $q=\cos^{-1}(-\alpha_J/4)$.
We also note that the peak at $q=0$ and dip at $q=\pi$ for $\alpha_J=1.0$ seen in Fig.~\ref{fig:spin_scf_frr}(b) can be assigned to the signature of dimer correlation.

The most important finding in this section is the fact that the chiral correlations arise at finite temperature from the dimer ground-state phase. This reminds us of similar behaviors seen in order-by-disorder phenomena, i.e., induced ordering at finite temperature from disordered ground state.~\cite{ANNNI,frustrated}  However, physics behind the present results and order by disorder looks different, because no disorder exists in the ground state in the present case.

%%%%%%%%%%%%%%%%%%%%%%%%%%%%%
\section{Conclusion}
We have investigated finite-temperature effects on vector chirality in spin-$\frac{1}{2}$ zigzag $XY$ chains, employing LT-DMRG.  We have compared two cases: with and without chiral LRO. The latter corresponds to the dimer phase. 

We have calculated the temperature dependence of dynamical and static correlations of vector chirality as well as static spin correlation. In the chiral phase, chiral LRO disappears at finite temperature as expected form the fact that there is no LRO at finite temperature in 1D systems with short-range interactions. In the dimer phase next to the chiral phase, static chiral correlation as well as spin correlation is enhanced with increasing temperature up to a characteristic temperature that is related to the gap magnitude of chiral excitation. This temperature-induced chiral correlation is a demonstration of the presence of a chiral state in excited states.

The present result also reveals usefulness of LT-DMRG for the numerical studies of 1D frustrated quantum spin systems at low temperature.

%%%%%%%%%%%%%%%%%%%%%%%%%%%%%
\acknowledgements
We thank J. Kokalj, P. Prelovsek, and S. Onoda for fruitful discussions.
This work was supported by Nanoscience Program of Next Generation Supercomputing Project, the Grant-in-Aid for Scientific Research (Grants No. 19052003 and No. 22340097) from MEXT, the Grant-in-Aid for the Global COE Program ``The Next Generation of Physics, Spun from Universality and Emergence,'' and the Yukawa International Program for Quark-Hadron Sciences at YITP, Kyoto University. A part of numerical calculations was performed in the supercomputing facilities in ISSP and ITC, the University of Tokyo, YITP and ACCMS, Kyoto University.  The financial support of JPSJ and MHEST under the Slovenia-Japan Research Cooperative Program is also acknowledged.

%\bibliography{string_prb,chirality}

\end{document}